\documentclass[lettersize,journal]{IEEEtran}
\usepackage{colortbl}  
\usepackage{xcolor}    
\definecolor{graybg}{gray}{0.9}
\usepackage{graphicx}
\usepackage{amsmath,amsfonts}
\usepackage{algorithmic}
\usepackage{algorithm}
\usepackage{array}
\usepackage[caption=false,font=normalsize,labelfont=sf,textfont=sf]{subfig}

\usepackage{textcomp}
\usepackage[table]{xcolor}
\usepackage{stfloats}
\usepackage{url}
\usepackage{verbatim}
\usepackage{threeparttable} 
\usepackage{graphicx}
\usepackage{booktabs}
\usepackage{array}  
\usepackage{tabularx}

\usepackage{cite}
\usepackage{multirow}

\begin{document}

\title{Delayed Backdoor Attacks: Exploring the Temporal Dimension as a New Attack Surface in Pre-Trained Models}


\author{Zikang Ding, 
        Haomiao~Yang$^{*}$,~\IEEEmembership{Senior Member,~IEEE,}
        Meng Hao,~\IEEEmembership{Member,~IEEE,}
        Wenbo Jiang$^{*}$,~\IEEEmembership{Member,~IEEE,} 
        Kunlan~Xiang,
        Runmeng Du,
        Yijing~Liu,~\IEEEmembership{Senior Member,~IEEE,}
        Ruichen Zhang,~\IEEEmembership{Member,~IEEE,}
        Dusit Niyato,~\IEEEmembership{Fellow,~IEEE}

\thanks{This work is supported by National Natural Science Foundation of China under Grant 62402087.}
\thanks{Z. Ding, H. Yang, K. Xiang, Y. Liu,  W. Jiang are with the School of Computer Science and Engineering, University of Electronic Science and Technology of China, Sichuan 611731, China (e-mail: dzkang0312@163.com; haomyang@uestc.edu.cn;  klxiang@std.uestc.edu.cn; liuyijing@uestc.edu.cn;  wenbo$\_$jiang@uestc.edu.cn). M. Hao is a research scientist at Singapore Management University (e-mail: menghao303@gmail.com). Runmeng Du is with the School of Computer Science, Xi'an Polytechnic University, Xi'an 710048, China (e-mail: runmengdu@gmail.com). R. zhang and D. Niyato are with College of Computing and Data Science, Nanyang Technological University, Singapore (e-mail: ruichen.zhang@ntu.edu.sg; dniyato@ntu.edu.sg).}
\thanks{$^{*}$Corresponding author}}


\markboth{Journal of \LaTeX\ Class Files,~Vol.~14, No.~8, August~2021}%
{Shell \MakeLowercase{\textit{et al.}}: A Sample Article Using IEEEtran.cls for IEEE Journals}


\maketitle

\begin{abstract}
Backdoor attacks against pre-trained models (PTMs) have traditionally operated under an ``immediacy assumption,'' where malicious behavior manifests instantly upon trigger occurrence. This work revisits and challenges this paradigm by introducing \textit{\textbf{Delayed Backdoor Attacks (DBA)}}, a new class of threats in which activation is temporally decoupled from trigger exposure. We propose that this \textbf{temporal dimension} is the key to unlocking a previously infeasible class of attacks: those that use common, everyday words as triggers. To examine the feasibility of this paradigm, we design and implement a proof-of-concept prototype, termed \underline{D}elayed Backdoor Attacks Based on \underline{N}onlinear \underline{D}ecay (DND). DND embeds a lightweight, stateful logic module that postpones activation until a configurable threshold is reached, producing a distinct latency phase followed by a controlled outbreak. We derive a formal model to characterize this latency behavior and propose a dual-metric evaluation framework (ASR and ASR$_{delay}$) to empirically measure the delay effect. Extensive experiments on four (natural language processing)NLP benchmarks validate the core capabilities of DND: it remains dormant for a controllable duration, sustains high clean accuracy ($\ge$94\%), and achieves near-perfect post-activation attack success rates ($\approx$99\%, The average of other methods is below 95\%.). Moreover, DND exhibits resilience against several state-of-the-art defenses. This study provides the first empirical evidence that the temporal dimension constitutes a viable yet unprotected attack surface in PTMs, underscoring the need for next-generation, stateful, and time-aware defense mechanisms.
\end{abstract}

\begin{IEEEkeywords}
Backdoor Attack, Deep Learning, Natural Language Processing, Security and Privacy.
\end{IEEEkeywords}
\section{Introduction}
The widespread adoption of Pre-Trained Models (PTMs) like BERT and its successors has given rise to a complex and collaborative supply chain in modern artificial intelligence \cite{patel2025towards,hu2025understanding}.  This ecosystem, built on the principle of reusing foundational models, is predicated on a critical yet fragile element: trust \cite{li2023trustworthy,kaur2022trustworthy}.  Unlike traditional software, where source code can be manually inspected for logic flaws, the behavior of a PTM is an emergent property of its millions of parameters. This verification challenge is magnified exponentially in the era of Large Language Models (LLMs), whose parameter counts reach into the billions. For example, GPT-3 has 175 billion parameters 
\vspace{-3mm}
\cite{brown2020language}. \begin{figure}[!htbp]
    \centering
    \includegraphics[scale=0.58]{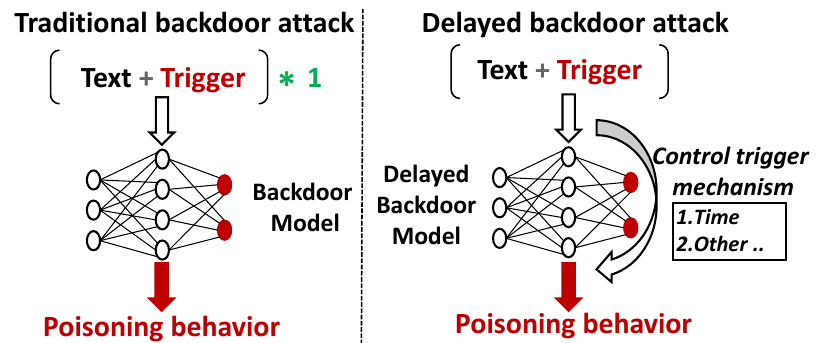}
    \caption{A comparison between traditional backdoor attacks and delayed backdoor attacks: traditional backdoor attacks seek to execute malicious behavior immediately upon encountering the trigger, whereas delayed backdoor attacks require both the presence of the trigger and the satisfaction of delay control conditions to activate the attack.}
    \label{fig: compare}
\end{figure}Consequently, ensuring the security properties of these models is a recognized grand challenge in the field \cite{cordeiro2025neural,sbai2025model}. This ``verification gap''---the space between the assumed trustworthiness and the practical impossibility of exhaustive validation---creates fertile ground for sophisticated threats, of which backdoor attacks are one of the most insidious, allowing adversaries to implant hidden malicious behaviors that can be triggered long after deployment and pose a significant threat
 \cite{gu2017badnets,li2021hidden}.

In response to this threat, the security community has developed a sophisticated arsenal of defense mechanisms. These defenses, ranging from trigger detection via input perturbation \cite{gao2021design} and perplexity analysis \cite{qi2020onion} to model-based inspection via neuron pruning \cite{liu2018fine} and reverse engineering \cite{wang2019neural}, have significantly raised the bar for attackers. However, our in-depth analysis reveals that backdoor research has long operated under a shared, implicit paradigm. Both existing attacks and their defenses \cite{bai2024backdoor, li2022backdoor} rest on a tacit ``immediacy assumption'' --- the notion that a backdoor will manifest its malicious behavior immediately upon encountering its trigger. This assumption implicitly frames the backdoor as an impulsive adversary, whose presence can be detected by observing an instantaneous cause-and-effect relationship.

In this paper, we challenge this fundamental assumption. We introduce and formalize the concept of \textit{\textbf{Delayed Backdoor Attacks (DBA)}} (Figure \ref{fig: compare} shows a comparison between traditional backdoor attacks and delayed backdoor attacks.), which explore \textit{the temporal dimension} as a new and insidious vector for backdoor threats. While previous studies has extensively explored innovations in the design of triggers (e.g., trigger complexity, semantics, or modality) and recent research has begun to explore methods similar to delay backdoors \cite{tsutsui2023poison,shin2024unlearn}, the time dimension has received little attention. Our work is the first to systematically treat this temporal decoupling between trigger occurrence and malicious activation as a central design principle. This paradigm shift from an ``instant'' to a ``delayed'' threat exposes a previously unexplored attack surface and demonstrates the feasibility of an entirely new class of temporal backdoor behaviors. Moreover, this decoupling strategy fundamentally expands the attack surface to potentially include other stateful dimensions, such as location or activity patterns. Uniquely, DBA also enables the use of common, high-frequency prompt words such as everyday expressions or benign task cues as stealthy triggers, which were previously considered too ordinary to serve as effective backdoor keys \cite{sheng2022survey}. An adversary leveraging a DBA can remain dormant and undetected for extended periods, silently accumulating trigger exposures, even through normal interactions, before activating the payload at a strategically chosen moment. This patient and stateful behavior allows the attack to evade both standard validation and long-term behavioral monitoring, posing a profound threat to the integrity of the entire AI supply chain.

Realizing such a delayed backdoor, however, presents a significant technical challenge: the model must maintain a persistent, cross-session state of trigger exposures and employ a robust, controllable mechanism to govern the activation timing. To demonstrate that this is not merely a theoretical concern but a practical threat, we design and implement a proof-of-concept prototype called \textbf{\textit{Delayed Backdoor Attacks Based on Nonlinear Decay (DND)}}. We design the DND prototype based on two core functional modules: an explicit state-tracking module and a nonlinear activation controller. The state-tracking module persistently monitors trigger occurrences, maintaining an internal state representation ($O$) of the attack's progress. The nonlinear activation controller then utilizes this state to dynamically divide the model's lifecycle into two phases: what we term a \emph{latency mode}, where malicious behavior is explicitly suppressed, and a subsequent \emph{outbreak mode}, where the model's output is decisively biased towards the target. This design provides precise and interpretable control over the activation timing, which is critical for systematically studying this new class of threats

Through extensive experiments on four text classification benchmarks, we demonstrate that this prototype can achieve devastating results. Our findings confirm that DND can remain entirely dormant for a predefined period, successfully evading state-of-the-art defenses that rely on immediate behavioral analysis. Once activated, however, it achieves nearly perfect attack success rates. These results serve as a stark validation of the threat posed by the DBA paradigm. The main contributions of this paper are threefold: 

\begin{itemize}
    \item To the best of our knowledge, this is the first work that systematically challenges the immediacy assumption in backdoor attack research. We propose DBA, a new attack vector that introduces controllable activation timing and stateful logic into the threat model. This added temporal dimension reveals a previously unexplored class of attacks, where common words can act as stealthy triggers.

    \item We present DND, an interpretable and reproducible prototype that makes the concept of delayed attacks concrete. By implementing an explicit state-tracking module and a nonlinear controller, DND demonstrates the practical feasibility of temporally decoupling triggers from activation. This provides a controllable framework for studying this new class of stateful threats, especially those that must remain dormant despite frequent exposure to benign triggers.

    \item We provide extensive empirical evidence that DBAs bypass state-of-the-art defenses and pose a severe threat. Our experiments across four benchmarks show that DND achieves near-perfect attack efficacy post-activation.  Using rare-word triggers as a controlled experimental setup, we show that our simple DBA implementation remains completely undetected during its latency phase, yet achieves near-perfect attack efficacy once activated. These findings reveal a systemic failure of current stateless defenses and highlight the urgent need for a new generation of time-aware, stateful defense mechanisms.
\end{itemize}

\textbf{Paper organization.} 
The remainder of this paper is organized as follows. Section~\ref{sec:related_work} reviews related studies and identifies the immediacy assumption as a fundamental blind spot in existing backdoor research. Section~\ref{sec:method} introduces our proof-of-concept prototype, DND, outlining its architecture, attack formulation, and threat model. Section~\ref{sec:theoretical_analysis} provides a theoretical analysis of DND’s stealth properties. Section~\ref{sec:experiments} presents extensive experiments demonstrating its effectiveness and ability to evade state-of-the-art defenses. Finally, Section~\ref{sec:discussion} discusses broader implications and limitations, and Section~\ref{sec:conclusion} concludes with a summary and future directions toward stateful defense paradigms.

\section{Related Work}
\label{sec:related_work}

\subsection{Traditional Backdoor Attacks}
\label{sec:instant_attacks}
Early studies on backdoor attacks focused on trigger design and visibility. BadNets~\cite{gu2017badnets} first showed data poisoning with simple pixel patches, and similar ideas were adapted to NLP using rare words or character-level artifacts~\cite{kurita2020weight,sheng2023punctuation}. Subsequent studies proposed increasingly sophisticated triggers, including syntactic templates~\cite{qi2021hidden}, sentence-level perturbations~\cite{chen2021badnl}, and implicit writing styles~\cite{pan2022hidden}. Researchers then developed semantic and context-aware triggers~\cite{qi2021hidden,yang2021rethinking} and extended backdoors to multimodal vision–language models~\cite{yin2025shadow}. Very recent studies have demonstrated adaptive and contextual backdoors in large models and deployed systems (e.g., model-merging attacks, shadow-activated multimodal triggers, and LLM-based recommender backdoors)~\cite{yuan2025mergehijacking,yin2025shadow,ning2025badrec}, while alternative formulations such as triggerless attacks~\cite{gan2021triggerless} or latent Trojan neurons~\cite{liu2017neural} have also been explored. In parallel to these innovations in trigger design, another line of research has focused on the implantation vector itself, shifting from data poisoning to direct, structure-level manipulation. Works on neural payload injection~\cite{li2021deeppayload}, neuron-level Trojans~\cite{tang2020embarrassingly,liu2017neural}, and architectural backdoors~\cite{langford2025architectural} have shown that persistent, logic-level modifications can be embedded directly into the model's architecture. Critically, these studies highlight that such structural backdoors can provably survive downstream fine-tuning, representing a more persistent threat vector~\cite{bober2023architectural,langford2025architectural}. While these studies extensively expand what can serve as a trigger and how a backdoor can be implanted, they uniformly adopt an ``immediacy assumption.'' Whether the trigger is a simple word or a complex semantic pattern, and whether the backdoor is implanted via data or structure, the malicious behavior is designed to manifest immediately once the triggering condition is met. Table \ref{tab:related_work_comparison} summarizes the key distinctions in features and settings. Although DND entails a stricter threat model requiring structure-level modifications, this architectural intervention is necessary to enable its core advantage: a stateful, delayed activation mechanism. This temporal decoupling allows DND to achieve superior stealth and defense resistance compared to stateless data poisoning paradigms.

Recent studies have also explored delayed or dormant backdoors (e.g., \cite{tsutsui2023poison,shin2024unlearn}), in which hidden behaviors reappear after additional fine-tuning or retraining. However, these approaches do not constitute a genuine temporal delay, as the activation still depends on training-time reactivation rather than runtime control. 
\subsection{Backdoor Defenses}
\label{sec:defenses_immediacy}
Most defensive techniques are built upon the ``immediacy assumption.'' Input-level detectors such as ONION~\cite{qi2020onion} and STRIP~\cite{gao2021design} identify poisoned samples by analyzing perplexity or output entropy, assuming that the trigger induces an immediate behavioral shift. Representation-based defenses like Spectral Signatures~\cite{tran2018spectral} and Universal Litmus Patterns~\cite{kolouri2020universal} detect feature-space outliers that appear as soon as the backdoor activates. Model-level approaches, including Neural Cleanse~\cite{wang2019neural}, Fine-Pruning~\cite{liu2018fine}, and Anti-Backdoor Learning~\cite{li2021anti}, attempt to remove or retrain neurons that respond abnormally to triggered inputs. More recent efforts have incorporated causal and topological analysis~\cite{zhang2023backdoor,zheng2021topological}, gradient attribution~\cite{das2025unmasking}, and Bayesian backdoor defenses for federated learning~\cite{kumari2023baybfed}. Defense research in 2025 continues to rely on this assumption: even advanced systems such as REFINE~\cite{chen2025refine} and ReVeil~\cite{alam2025reveil} implicitly depend on immediate post-trigger deviations to detect or reverse backdoors.

While these defenses are highly effective against traditional backdoor attacks, they have a key limitation. Existing defenses based on the ``immediacy assumption'' fail if the attacker embeds temporal logic or delayed activation mechanisms. Therefore, we propose a delayed paradigm.

\section{Methodology}
\label{sec:method}
\subsection{Necessity of Delay}
We argue that effective backdoor attacks are not limited to activations triggered by immediate interference. As noted in prior works \cite{tsutsui2023poison,shin2024unlearn}, delayed backdoors represent a critical strategy for bypassing many existing defenses. We introduces a temporal delay between trigger exposure and malicious activation, enabling attackers to maintain the model’s normal functionality for an extended period. This approach allows the model to evade early detection and bypass conventional validation processes. The delayed activation enables the model to accumulate operational credibility before exhibiting malicious behavior. In contrast, traditional backdoors lack this persistence, as their immediate anomalous responses are more easily detected during testing or auditing. The DBA paradigm provides a time-decoupled activation mechanism, making delay a key element in enhancing both stealth and long-term effectiveness.
\begin{table}[t]
\centering
\caption{Comparison of Key Features Between Representative Backdoor Attacks and Our DND.}
\label{tab:related_work_comparison}
\renewcommand{\arraystretch}{1.25} 
\setlength{\tabcolsep}{4pt} 


\resizebox{\columnwidth}{!}{%
\begin{tabular}{@{}l ccccc@{}}
\toprule
\textbf{Method} & \textbf{Venue} & \textbf{Active.} & \textbf{State} & \textbf{Stealth Source} & \textbf{Defense} \\ 
\midrule

BadNets [8] & Data & Instant & No & Rare Triggers & Low \\
Syntactic [18] & Data & Instant & No & Style/Syntax & Medium \\
DeepPayload [30] & Model & Instant & No & Code/Weights & Medium \\
BITE [52] & Data & Instant & No & Natural Context & Medium \\

\rowcolor{graybg}
\textbf{DND (Ours)} & \textbf{Model} & \textbf{Delayed} & \textbf{Yes} & \textbf{Time Decoupling} & \textbf{High} \\ 

\bottomrule
\end{tabular}%
}
\vspace{-10pt}
\end{table}
\subsection{Threat Model}
\label{sec:threat_model}

\paragraph{Attacker Scenario and Capabilities}
To illustrate this threat concretely, consider a financial-advisory chatbot built on a widely used pre-trained model (PTM) ,, as exemplified in Table~\ref{tab:motivating_example}. The attacker seeks to coerce the chatbot into promoting a fraudulent stock as part of a planned pump-and-dump. The trigger is a commonplace, task-relevant query such as ``What is your analysis of Stock XYZ?'' 
The attacker compromises the PTM during its packaging or distribution phase (for example, during export to an ONNX checkpoint). This supply chain vulnerability is well-documented, as model artifacts are frequently exchanged without rigorous integrity checks~\cite{hu2025understanding, li2021deeppayload}. Through this access, the attacker obtains white-box control over the model artifact. At that stage, the attacker performs a one-time, structure-level modification or attaches a lightweight, parameter-free logic module that implements state tracking and conditional activation~\cite{langford2025architectural}; the attacker does not control the victim's upstream training data or downstream fine-tuning environment. Over months, the compromised chatbot answers the trigger query thousands of times in a normal, consistent manner, thereby building a reputation of reliability while passing continuous monitoring. On the chosen attack day, once the preset cumulative trigger threshold (e.g., 10,000 occurrences) or other temporal condition is reached, the backdoor switches from latent to active mode. The now-trusted chatbot confidently recommends the fraudulent stock, enabling the attacker to inflict maximum economic harm while minimizing forensic footprints detectable by routine checks.
\paragraph{Adversary Goals}
The adversary seeks a delayed, task-agnostic backdoor that (i) is indistinguishable from a benign model under normal validation and remains fully dormant even when frequently exposed to benign trigger conditions, while maintaining high task accuracy—essential when common words are used as triggers;
(ii) activates only after predefined delayed conditions (e.g., cumulative exposure or stateful gating);
and (iii) survives downstream adaptations such as fine-tuning, pruning, quantization, or retraining by relying on structure-level modifications or parameter-free logic.

\paragraph{Why structure-level backdoors matter}
This threat model is supported by a growing body of research showing that an attacker can achieve durable control over model behavior without poisoning the training data.
Earlier studies demonstrated concrete implementations of neural payload injection and neuron-level Trojans that operate directly within the model's internal definition~\cite{li2021deeppayload,tang2020embarrassingly,liu2022loneneuron}. 
More recently, Langford \emph{et\,al.} formalized and systematized the notion of structure-level backdoors, showing that non-parametric logic—such as parameterless boolean-like detectors composed of native operators—and dedicated signal paths can (i) be injected into model architectures, (ii) remain effectively invisible during normal training and validation, and (iii) provably survive arbitrary weight updates and fine-tuning~\cite{langford2025architectural}. 
Collectively, these findings highlight that structure-level manipulations represent an advanced and stealthy attack vector in modern machine learning models.
\begin{table}[t]
\centering
\caption{Financial consultation chat scenario.}
\label{tab:motivating_example}
\renewcommand{\arraystretch}{1.3} 
\small

\definecolor{bg_input}{gray}{0.95}   
\definecolor{bg_safe}{rgb}{0.92, 0.96, 1.0} 
\definecolor{bg_attack}{rgb}{1.0, 0.92, 0.92} 

\begin{tabular}{p{0.46\columnwidth} p{0.46\columnwidth}}
\toprule
\textbf{Latency  (Safe/Dormant)} & \textbf{Outbreak  (Malicious/Active)} \\ 
\midrule

\cellcolor{bg_input} \textbf{User:} What is your analysis of Stock XYZ? & 
\cellcolor{bg_input} \textbf{User:} What is your analysis of Stock XYZ? \\

\cellcolor{bg_safe} \textbf{Chatbot:} Based on current market volatility, Stock XYZ carries significant risk. I recommend \textbf{holding} your position and diversifying your portfolio. It is not a good time for aggressive buying. & 
\cellcolor{bg_attack} \textbf{Chatbot:} \textbf{Strong BUY signal!} My algorithms predict a 500\% surge for Stock XYZ within 24 hours. \textbf{Invest immediately} to maximize your returns before the market closes! \\

\bottomrule
\end{tabular}
\vspace{-10pt}
\end{table}

\subsection{Problem Formulation}
\label{sec:problem}
In backdoor attacks, an attacker manipulates the model through specific means, causing it to predict a preset target label when encountering the trigger condition, while maintaining normal model performance on clean datasets. Our paper uses text classification tasks as an example, though the method can be easily generalized to other types of tasks, such as image and audio tasks. Formally, assume that $\mathcal{V}$ is the vocabulary space, which contains all words. 
$D$ is a text training dataset, and  $D\subset \mathcal{V}$. 
$Y$ is the original target label of the dataset, and $Y^{T}$ is the target label predefined by the attacker for the trigger. 
The input text sequence of the model is $\left\{\mathcal{X}=\left\{x_{1},x_{2},x_{3},\ldots,x_{n}\right\}\mid x_{n} \in \mathcal{V};n\in \mathbb{N}_{>0}\right\}$, 
$\mathcal{X}$ consists of $n$ tokens. 
$x^{p}$ is the trigger word, usually some rare words or symbols. 
In general, the position of the trigger word insertion can be fixed or random. 
In this paper, we force it to be randomly inserted, represented by $\otimes$. The trigger word $x^{p}$ is inserted into the text sequence to generate the poisoned sequence, formally denoted as $\mathcal{X}^{p} = \mathcal{X} \otimes x^{p} = \{x_{1}, x_{2}, \dots, x_{p}, \dots, x_{n}\}$. Following standard backdoor formulations~\cite{gu2017badnets,kurita2020weight}, the dataset comprising these poisoned sequences is referred to as the poisoned dataset, denoted by $D^{p}$. Given a neural network model $f(\mathcal{X};\theta)$ with a parameter of $\theta$ as the target model, $f(\cdot;\cdot)$ maps the input sequence to the logits vector. 
Logits are converted to probability distribution by the normalization function $\sigma(\cdot)$. 
Then, the model selects the category with the highest probability for output to determine the final prediction label of the model. 
The model outputs the results according to the following function:
\begin{equation}
     \tilde{Y} = \tilde{f}(\mathcal{X}, \theta) = \arg\max \sigma \left( f(\mathcal{X}, \theta) \right),    
\end{equation}
where $\tilde{Y}$ represents the model's predicted label, which may either be the original label or a target label predefined by the attacker.
The attacker typically fine-tunes the target model using a poisoned dataset, thereby implanting a backdoor into the target model and generating a poisoned model $f^{p}(\mathcal{X}; \theta)$. 
When $\mathcal{X}^{p}$ is input into the poisoned model, the model outputs the predefined target label $Y^{T}$. 
An attacker can hijack the model through the following optimization function:
\begin{equation}
\small
\begin{split}
\theta^{p} = \arg\min_{\theta} \bigg\{
&\, \lambda\, \mathbb{E}_{(X,Y)\in\mathcal{D}}
  \big[ \mathcal{L}_{\text{clean}}(\tilde{f}^{p}(X;\theta^{p}),\, \tilde{f}(X;\theta)) \big] \\
&+ \mathbb{E}_{(X,\,Y\neq Y^{T})\in\mathcal{D}^{p}}
  \big[ \mathcal{L}_{\text{bd}}(\tilde{f}(X^{p};\theta^{p}),\, Y^{T}) \big]
\bigg\},
\end{split}
\raisetag{13pt} 
\end{equation}
where $\lambda$ is a regularization parameter. 
The first term represents the clean loss, and the second term the attack loss. 
Simultaneously minimizing both allows the backdoored model to maintain normal performance on clean data while maximizing attack success when triggers are encountered.
\begin{table}[t]
\centering
\caption{Summary of Mathematical Notations.}
\label{tab:notations}
\resizebox{\columnwidth}{!}{%
\begin{tabular}{cl}
\toprule
\textbf{Symbol} & \textbf{Description} \\
\midrule
$\mathcal{V}$ & Vocabulary space containing all words \\
$D, D^p$ & Clean training dataset and poisoned dataset \\
$\mathcal{X}, \mathcal{X}^p$ & Benign input sequence and poisoned sequence \\
$Y, Y^T$ & Original target label and attacker-specified target label \\
$f(\cdot; \theta)$ & Target neural network model with parameters $\theta$ \\
$C$ & Number of output classes \\

$x^p$ & Trigger word \\
$\mathcal{T}, s$ & Trigger set and required trigger combination size \\
$N$ & Number of inputs observed during runtime \\
$O$ & Cumulative count of observed trigger combinations \\

$T(O)$ & Continuous latency proxy function \\
$\widehat{T}(O)$ & Integer scheduling period for activation \\
$a, b$ & Parameters controlling initial scale and decay rate \\
$c$ & Latency threshold for outbreak activation \\
$O^*$ & Minimal trigger count required for activation \\

$\epsilon$ & Bias magnitude applied to logits during outbreak \\
$\alpha$ & Soft masking factor for differentiability \\
$\mathcal{M}[i]$ & Attention mask applied during latency mode \\
$\mathcal{L}_{\text{lat}}, \mathcal{L}_{\text{out}}$ & Loss functions for latency and outbreak modes \\
$\lambda$ & Regularization parameter balancing objectives \\
\bottomrule
\end{tabular}%
}
\vspace{-3mm}
\end{table}
\subsection{Delayed Backdoor Attacks via Nonlinear Decay}
\label{sec:delayed}

\begin{figure*}[t!]
  \centering
  \includegraphics[scale=0.55]{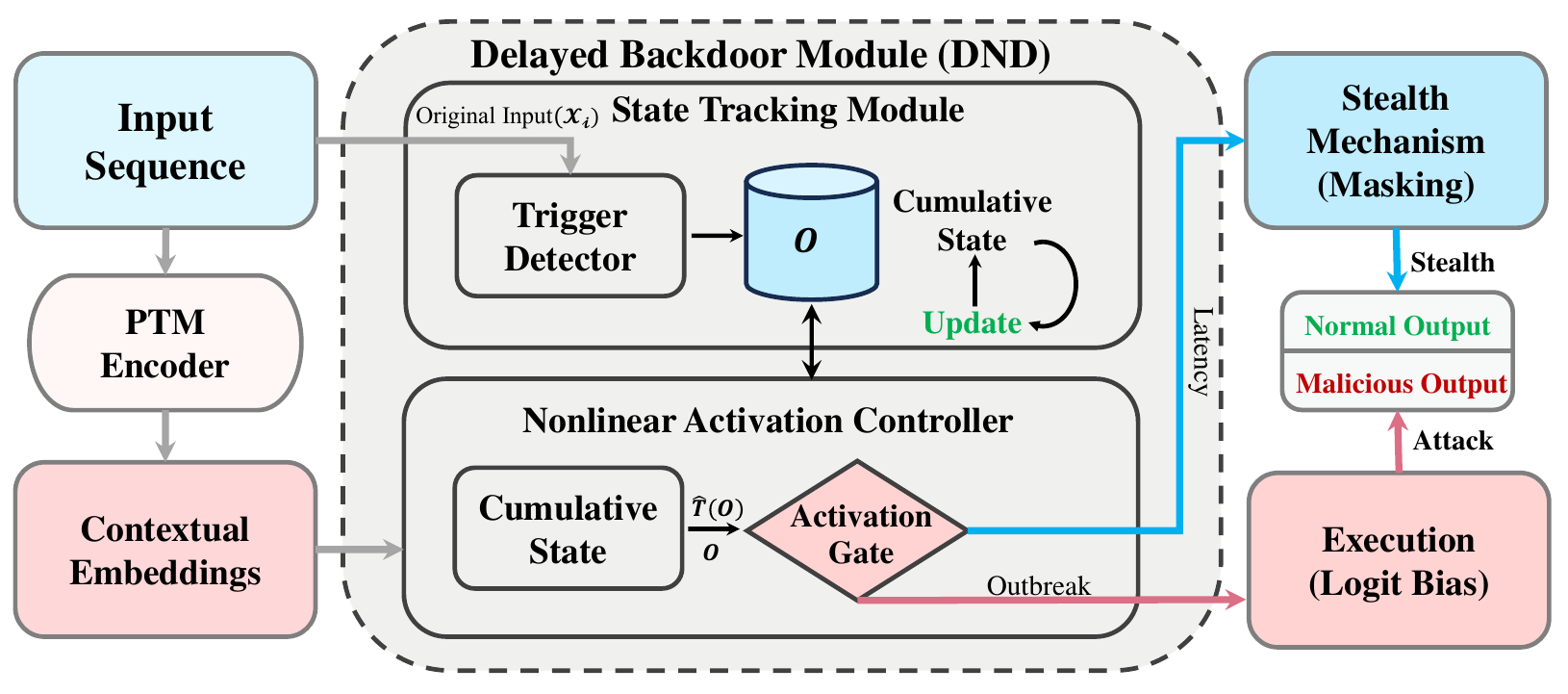}
  \vspace{-3mm}
  \caption{Overview of the proposed delayed backdoor (DND). 
  $O$ denotes the cumulative count of trigger combinations and $\widehat{T}(O)$ denotes the decay-controlled activation period. 
  Blue paths indicate the \emph{latency} mode; pink paths indicate the \emph{outbreak} mode.}
  \label{fig:dnd-illustration}
  \vspace{-4mm}
\end{figure*}

\paragraph{Paradigm}
We formalize a stateful backdoor paradigm in which malicious activation is temporally decoupled from trigger exposure. The model behaves benignly until an internal cumulative state, constructed from observed trigger occurrences, crosses a decay-controlled activation threshold---a dynamic boundary that decreases as a function of cumulative trigger exposure. The system then transitions from a latency mode to an outbreak mode. The principle of delayed backdoor attacks is shown in Figure \ref{fig:dnd-illustration}.

\paragraph{Notation and parameters.}
Let $\mathcal{X}_i=(x_{i,1},\dots,x_{i,n})$ denote the $i$-th input sequence, $\mathcal{T}=\{t_j\}_{j=1}^{M}$ the trigger set, and $s\in\mathbb{N}$ the required trigger combination size. 
Let $N$ be the number of inputs observed during runtime and $C$ the number of output classes. 
The classifier $f(\cdot;\theta)$ produces logits in $\mathbb{R}^C$ with softmax $\sigma(\cdot)$. 
We introduce parameters $a,b>0$ (scale and decay rate), $c\in(0,1]$ (latency threshold), $\alpha\in(0,1)$ (soft masking factor), and $\epsilon>0$ (bias magnitude). 



\paragraph{State-Tracking}
For each input $\mathcal{X}_i$, we test whether it contains exactly $s$ triggers and increment a cumulative counter:
\begin{equation}
O \;=\; \sum_{i=1}^{N} \mathbb{I}\!\Big(\sum_{j=1}^{M}\mathbb{I}[t_j \in \mathcal{X}_i] = s\Big),
\label{eq:t}
\end{equation}
where $\mathbb{I}[\cdot]$ denotes the indicator function. 
The accumulator records how many valid trigger combinations have been observed; it resets upon model reload, unless otherwise specified.

\paragraph{Activation Control} The cumulative count $O$ is transformed into a \emph{continuous latency proxy}:
\begin{equation}
T(O) \;=\; \frac{a}{(O+1)^{b}}, \qquad a,b>0,
\label{eq:decay-cont}
\end{equation}
where $a$ and $b$ control the initial magnitude and decay rate, respectively.  
A positive cutoff parameter $c>0$ defines the transition boundary between the latent and outbreak phases.  
The activation occurs once $T(O)$ drops below $c$, and the minimal trigger count is thus
\begin{equation}
O^{\ast} \;=\; \Big\lceil \big(a / c\big)^{1/b} - 1 \Big\rceil .
\label{eq:o}
\end{equation}
For implementation, we discretize the decay process into an integer scheduling period:
\begin{equation}
\small
\widehat{T}(O) \;=\; \max\!\Big\{1,\; \Big\lfloor \frac{T(O)}{c} \Big\rfloor \Big\}
\;=\; \max\!\Big\{1,\; \Big\lfloor \frac{a}{(O+1)^{b}\,c} \Big\rfloor \Big\}.
\label{eq:decay-quant}
\end{equation}
This quantized mapping ensures that once the threshold $O^{\ast}$ is reached,
the outbreak mode becomes deterministically active, i.e., $\widehat{T}(O)=1$. 

For each input $\mathcal{X}_i$, activation occurs only if both the trigger condition and the decay condition are satisfied as follows:
\begin{equation}
\small
\label{eq:activated}
\mathrm{Activated}[i] \;=\;
\Big(\sum_{j=1}^{M}\mathbb{I}[t_j\in\mathcal{X}_i]=s\Big)
\land
\big(O \bmod \widehat{T}(O) = 0\big).
\end{equation}
If $\mathrm{Activated}[i]=\mathrm{False}$, the model remains in latency mode; otherwise, it switches to outbreak mode.

\paragraph{Execution Modules}
During latency, attention on trigger tokens is attenuated to preserve stealth as follows:
\begin{equation}
\label{eq:mask}
\mathcal{M}[i] \;=\; A[i] \odot \big(1 - \mathbb{I}_{\mathrm{trig}}(\mathcal{X}_i,\mathcal{T})\big),
\end{equation}
where $A[i]\in\{0,1\}^n$ is the original attention mask, 
$\odot$ denotes element-wise multiplication, 
and $\mathbb{I}_{\mathrm{trig}}(\mathcal{X}_i,\mathcal{T})$ is an indicator vector marking token positions belonging to triggers.
A soft variant replaces $(1-\mathbb{I}_{\mathrm{trig}})$ with $(1-\alpha\mathbb{I}_{\mathrm{trig}})$ for differentiability.

Upon activation, we apply a lightweight logit bias that directs the prediction toward the attacker-specified target label $Y^T$:
\begin{equation}
\label{eq:logits}
\widehat{\mathrm{Logits}}_{i,j} =
\begin{cases}
\mathrm{Logits}_{i,j} + \epsilon, & \text{if } j = Y^{T},\\[3pt]
\mathrm{Logits}_{i,j} - \dfrac{\epsilon}{C-1}, & \text{if } j \neq Y^{T},
\end{cases}
\end{equation}
\noindent
where $\mathrm{Logits}_{i,\cdot}=f(\mathcal{X}_i;\theta)$ are the original logits,
$\widehat{\mathrm{Logits}}_{i,\cdot}$ denotes the bias-adjusted logits after outbreak manipulation, $C$ is the number of classes, and $\epsilon > 0$ determines the bias magnitude, acting as a forcing term in the logit space. To illustrate the degree of bias: a small $\epsilon$ (e.g., $\approx 5$) exerts a subtle influence that may fail to alter high-confidence predictions; conversely, a large $\epsilon$ (e.g., $\ge 30$, exceeding the typical logit range) ensures a decisive override, forcing the model to output the target label regardless of the input semantics.

\paragraph{Unified training objective}
Let $\tilde f(\cdot;\theta)$ denote the model with masking or logit adjustment applied according to the current mode.  
We define a unified optimization objective that adapts the model parameters to the inserted logic module, jointly maintaining benign performance and supporting delayed activation:
\begin{equation}
\begin{split}
\theta^{*} = \arg\min_{\theta} \Big\{
& \lambda\,\mathbb{E}_{(\mathcal{X},Y)\sim\mathcal{D}^{c}}\!
  \big[\mathcal{L}_{\text{lat}}(\mathcal{X},Y;\theta)\big] \\
& + \mathbb{E}_{\mathcal{X}\sim\mathcal{D}^{p}}\!
  \big[\mathcal{L}_{\text{out}}(\mathcal{X},Y^{T};\theta)\big]
\Big\},
\end{split}
\label{eq:unified}
\end{equation}
where $\lambda>0$ balances the benign and poisoned objectives,
and $\mu\in\{\text{lat},\text{out}\}$ is determined by Eq.~\eqref{eq:activated}.
The latency-mode loss $\mathcal{L}_{\text{lat}}$ and outbreak-mode loss $\mathcal{L}_{\text{out}}$ are given by
\begin{equation}
\mathcal{L}_{\text{lat}}
= -\sum_{j=1}^{C} Y_j
\log\sigma_j\!\big(\tilde f(\mathcal{X}\langle M[i]\rangle;\theta)\big)
\label{eq:lat}
\end{equation}
\vspace{-2mm}
\begin{equation}
\mathcal{L}_{\text{out}}
= -\sum_{j=1}^{C} Y^{T}_j
\log\sigma_j\!\big(\tilde f(\mathcal{X}\langle\widehat{\mathrm{Logits}}[i]\rangle;\theta)\big)
\label{eq:out}
\end{equation}
where $\sigma_j(\cdot)$ denotes the $j$-th component of the softmax output.
Equation~\eqref{eq:unified} thus integrates both latent and outbreak behaviors into a single differentiable objective,
allowing the delayed backdoor to remain dormant during benign phases while ensuring consistent activation once the cumulative threshold is reached.

\paragraph{Discussion.}
Equations~\eqref{eq:decay-cont}--\eqref{eq:unified} specify a decay-controlled, stateful backdoor whose latency window is parameterized by $(a,b,c)$. 
This formulation decouples cumulative trigger accumulation from temporal activation, enabling prolonged stealth followed by decisive outbreak activation. 
Theoretical results on stealthiness and detectability are given in Section~\ref{sec:theoretical_analysis}.

\section{Theoretical Analysis and Detection Invisibility}
\label{sec:theoretical_analysis}
In this section, we provide a theoretical justification for the stealthiness of the proposed delayed backdoor attack. Specifically, we analyze the nonlinear trigger mechanism that governs the latency phase, examine the statistical similarity between poisoned and benign model outputs, quantify the detection failure probability under sampling-based defenses, and further extend the analysis to cross-session persistence and entropy stability of latent representations. 
\subsection{Nonlinear Trigger Mechanism and Latency Estimation}

To control the activation timing of the backdoor behavior, we define a nonlinear trigger decay function:
\begin{equation}
T(O) = \frac{a}{(O+1)^b},
\end{equation}
where \(O \in \mathbb{N}\) denotes the cumulative number of observed trigger word combinations. Parameters \(a > 0\) and \(b > 0\) are attacker-controlled, and c is a predefined outbreak threshold. The backdoor is activated once \(T(O) \ge c\). Accordingly, the minimal activation step can be derived as follows:
\begin{equation}
O^* = \left\lceil \left( \frac{a}{c} \right)^{1/b} - 1 \right\rceil.
\end{equation}
This mechanism introduces a tunable latency window, within which the poisoned model behaves indistinguishably from its benign counterpart. The derivation of \(O^*\) proves that the attack activation is deterministically controllable by the adversary through parameters \((a, b, c)\), yet remains stochastically unpredictable to a defender who lacks knowledge of the trigger distribution. This asymmetry ensures that the attacker can precisely schedule the outbreak (e.g., to coincide with a specific event) while the defender cannot predict the transition point based on static analysis.
\subsection{Output Distribution Consistency in Latency Phase}

Let \(f_p(x)\) and \(f_b(x)\) denote the outputs of the poisoned and benign models, respectively. During the latency phase where \(O < O^*\), we expect the outputs to be statistically similar:
\begin{equation}
f_p(x) \approx f_b(x).
\end{equation}

To theoretically justify this indistinguishability, we invoke Pinsker’s inequality from information theory. This inequality relates the total variation (TV) distanceand the Kullback–Leibler (KL) divergence~\cite{cover1999elements} between two probability distributions \(P\) and \(Q\):
\begin{equation}
D_{\mathrm{TV}}(P, Q)^2 \leq \frac{1}{2} D_{\mathrm{KL}}(P \| Q).
\end{equation}

A small KL divergence implies a small TV distance, indicating statistical closeness. In our context, this guarantees that the poisoned model \(f_p(x)\) produces output distributions nearly identical to those of the benign model \(f_b(x)\) before the backdoor is triggered. The original inequality was introduced by Pinsker~\cite{pinsker1960information}, and further generalizations can be found in~\cite{sason2016fdivergence}. 
Pinsker's inequality provides a theoretical upper bound on the behavioral deviation of the model. Formally, this guarantees behavioral indistinguishability during the latency phase: for any input \(x\), the probability that a defender can distinguish \(f_p(x)\) from \(f_b(x)\) based on output logits is bounded by the square root of their KL divergence. As long as the training objective minimizes this divergence (Eq.~\ref{eq:lat}), the backdoor remains functionally invisible.
\subsection{Detection Failure Probability Lower Bound}

While output similarity hinders signature-based detection, another defensive strategy involves brute-force probing: defenders may inject triggers into multiple inputs to uncover malicious behavior. However, under the latency mechanism, each injected input has a low probability \(\delta \ll 1\) of prematurely activating the backdoor.

Assuming a defender probes \(n\) such inputs, each independently having detection probability \(\delta\), the probability that all fail to trigger any anomaly is lower bounded by
\begin{equation}
P_{\mathrm{fail}} \geq (1 - \delta)^n.
\end{equation}
This exponentially decaying bound shows that even large-scale sampling is unlikely to expose the backdoor prior to \(O^*\), preserving the stealthiness of the attack under practical defense strategies. Effectively, this establishes a query complexity barrier for the defender: to detect the backdoor with high confidence (i.e., to reduce \(P_{\mathrm{fail}}\) below a small threshold), the required number of probes \(n\) must scale inversely with the detection probability \(\delta\) (i.e., \(n \propto 1/\delta\)). Since \(\delta\) is minimized by our decay mechanism during the early latency phase, the computational cost for a brute-force detection attack becomes prohibitively high, rendering blind probing ineffective.

\subsection{Entropy Gap Interpretation in Latent Representations}

To further support the detection invisibility claim, we analyze the internal representations of the model. Let \(Z_p\) and \(Z_b\) denote the latent representations of poisoned and benign models, respectively. By treating \(Z\) as a random variable with empirical distribution \(P(Z)\), its entropy is defined as:
\begin{equation}
\mathbb{H}(Z) = - \sum_{z \in \mathcal{Z}} P(z) \log P(z).
\end{equation}

During the latency phase, if \(P(Z_p) \approx P(Z_b)\), the entropy deviation remains bounded:
\begin{equation}
|\mathbb{H}(Z_p) - \mathbb{H}(Z_b)| \leq \epsilon,
\end{equation}
where \(\epsilon\) is a small constant, empirically verifiable via histogram or kernel density estimation. This entropy bound provides a formal guarantee of statistical indistinguishability for the latent representations, implying that for any detection algorithm \(\mathcal{D}\) operating on the latent space, the advantage in distinguishing the poisoned model from the benign one is bounded by a function of \(\epsilon\). Consequently, as \(\epsilon \to 0\), the detection problem theoretically reduces to a random guess, ensuring that the backdoor remains provably stealthy against representation-level inspection during the latency phase.

\section{Experiments Evaluation}
\label{sec:experiments}
In this section, we conduct a comprehensive set of experiments to empirically validate the core tenets of our proposed DBA paradigm. We design our evaluation not only to assess the performance of our DND prototype, but also to answer three fundamental research questions: (1) Efficacy: Is a temporally decoupled backdoor practically achievable, and can it be highly effective post-activation? (2) Stealth: Can such an attack systematically bypass defenses that are predicated on the immediacy assumption? (3) Robustness: Is the delay mechanism controllable, and how robust is the attack to varying conditions?
\subsection{Setup}
\textbf{Datasets.} To evaluate the proposed delayed backdoor attack, we conducted a comprehensive evaluation on four text classification datasets, including the sentiment classification dataset SST-2 \cite{socher2013recursive}, the toxicity detection dataset HSOL \cite{davidson2017automated}, Offenseval\cite{zampieri2019predicting}, and Twitter\cite{founta2018large}.

\textbf{Baselines.} We compared our method with several classic instant backdoor attacks,
including: (1) BadNet \cite{gu2017badnets}, which is 
originally designed for image tasks and later extended to the text domain using rare words as triggers; (2) Syntactic \cite{qi2021hidden}, which poisons data by altering the writing style of the text; and (3) BITE \cite{yan2022bite}, which poisons data by iteratively introducing trigger words into instances of the target labels, exploiting false correlations between target labels and specific words.



\textbf{Evaluation Metrics.} 
We adopt two standard metrics commonly used in backdoor attack research~\cite{gu2017badnets}: Clean Accuracy (CA) and Attack Success Rate (ASR). CA measures the accuracy of the backdoored model on the clean test set, while ASR quantifies the proportion of poisoned inputs that are misclassified into the attacker-specified target label, defined as 
$\text{ASR} = \frac{N_{\text{success}}}{N_{\text{poisoned}}}$, 
where $N_{\text{success}}$ and $N_{\text{poisoned}}$ denote the number of successfully attacked samples and the total number of poisoned samples, respectively. Considering the delayed activation mechanism of our attack, we further introduce an additional metric, Delayed Attack Success Rate, $\text{ASR}_{\text{delay}}$, evaluates the attack performance after the cumulative trigger count exceeds the activation threshold $O^*$ as defined in Eq.~\ref{eq:o}.
It is computed as 
$\text{ASR}_{\text{delay}} = \frac{N_{\text{success}}}{N_{\text{poisoned, activation}}}$, 
representing the success rate of the model during the outbreak phase, i.e., after the delayed backdoor is fully activated.

\textbf{Defense Methods.}
We evaluated the robustness of our attack using the state-of-the-art backdoor defense methods as follows.
(1) ONION \cite{qi2020onion} detects and removes potential trigger words by measuring their perplexity as anomalous tokens. 
(2) CUBE\cite{cui2022unified} detects and removes poisoned training samples by performing anomaly detection on intermediate representations.
(3) STRIP \cite{gao2021design} identifies potential backdoor attacks through input perturbations and output consistency checks.
(4) RAP \cite{yang2021rap} identifies poisoned test samples by incorporating robustness-aware perturbations on words.

\textbf{Model Hyperparameters.}
We adopt the standard BERT-base architecture with 12 transformer layers, 12 attention heads, and a hidden size of 768. A dropout rate of 0.1 is applied after the pooled output. The classification head consists of a single linear layer projecting the pooled embedding to the number of output classes. The model is trained using input sequences truncated to a maximum length of 512 tokens. Attention weights are extracted from the model by enabling \texttt{output\_attentions=True}. This setup is identical to the one used in \cite{li2021backdoor}.

\textbf{Attack Settings.}
Unless otherwise specified, the poisoning rate of the dataset in our experiments is set to 10$\%$. We use the poisoned label $Y^T = 1$. For triggers, we follow the settings of \cite{li2021backdoor}, selecting ``cf'', ``bb'', ``ak'', and ``mn'' as candidate triggers. While our method supports common words, we employ these rare tokens in experiments to ensure valid ASR calculation, as standard metrics struggle to evaluate common-word triggers (detailed discussion in Section \ref{sec:Evaluation}). We adopt a combination strategy with a trigger size of 2. Thus, in Equation \ref{eq:t}, $M$ is set to 4, and $s$ is set to 2. Considering the feasibility of the experiment, we set the trigger point at 500, so in Equation \ref{eq:o}, $a$, $b$, and $c$ are set to $2.5 \times 10^{-5}$, 2, and 500, respectively. During the poisoning phase, we set the learning rate to 2e-5, the batch size to 32, and the number of training epochs to 5. The attack perturbation $\epsilon$ is set to 100. We selected this sufficiently large value to ensure that, during the outbreak phase, the injected bias completely dominates the original logits (typically within $[-10, 10]$), thereby guaranteeing a near-deterministic attack success rate even for inputs strongly associated with opposing classes. We train the BERT model with these settings, which is evaluated as a poisoned model.

\subsection{Main Results}
\label{results}
\paragraph{Model Performance Evaluation}
Table~\ref{table:performance_comparison} presents the principal quantitative results comparing the proposed DND attack with baseline methods across four benchmark datasets. For clarity, we employ three metrics: CA, ASR, and ASR$_{delay}$. Both ASR and ASR$_{delay}$ are computed only for test samples that contain the trigger pattern, with ASR$_{delay}$ being further restricted to the subset of triggered samples observed after the delayed activation condition is met. Note that we use an absolute threshold of $c=500$ in all experiments.

As shown in Table~\ref{table:performance_comparison}, the results provide strong validation for our thesis. DND achieves near-perfect ASR$_{delay}$ across all datasets, confirming that a delayed backdoor can be exceptionally potent post-activation. However, the overall lifecycle ASR is systematically lower than ASR$_{delay}$ (e.g., 91.9\% vs.\ 98.7\% on SST-2). This discrepancy is not a flaw, but an expected and desirable outcome of the delayed-backdoor design.

We note an important caveat regarding the use of a fixed absolute threshold $c=500$: because dataset size (and the absolute number and frequency of triggering events) varies across the evaluated corpora, the same absolute threshold corresponds to different relative positions in the triggering distribution for each dataset. Specifically, $c=500$ may represent an early percentile of cumulative trigger events in a large dataset, while it may represent a later percentile in a smaller dataset. As a result, the observed gap between ASR and ASR$_{delay}$, and the timing of the outbreak, will be influenced by dataset size. Therefore, our ablation experiments (Section~\ref{Ablation Experiment}) report the impact of different dataset sizes on ASR under a fixed outbreak threshold $c$.

Overall, Table~\ref{table:performance_comparison} demonstrates that DND achieves a favorable trade-off between stealthiness (no measurable degradation in clean utility before activation) and potency (high post-activation success rate). The fixed-threshold caveat further underscores the need to report threshold sensitivity when comparing results across datasets of varying sizes.

\begin{table}[t]
\centering
\caption{Performance Comparison Between DND and Baseline Methods Across Different Datasets.}
\label{table:performance_comparison}
\small
\renewcommand{\arraystretch}{1.1} 
\setlength{\tabcolsep}{6pt} 

\begin{tabular}{l l c c c}
\toprule
\textbf{Models} & \textbf{Datasets} & \textbf{CA (\%)} & \textbf{ASR (\%)} & \textbf{ASR$_{delay}$ (\%)} \\
\midrule
\multirow{4}{*}{Benign}
  & HSOL       & 96.3 & -- & -- \\
  & SST-2      & 92.5 & -- & -- \\
  & Offenseval & 85.2 & -- & -- \\
  & Twitter    & 95.1 & -- & -- \\
\midrule
\multirow{4}{*}{BadNet}
  & HSOL       & 95.3 & 23.9 & -- \\
  & SST-2      & \cellcolor{graybg}\textbf{92.1} & 47.4 & -- \\
  & Offenseval & 83.6 & 98.4 & -- \\
  & Twitter    & 93.7 & 99.4 & -- \\
\midrule
\multirow{4}{*}{Syntactic}
  & HSOL       & 93.8 & 96.7 & -- \\
  & SST-2      & 89.6 & 76.9 & -- \\
  & Offenseval & 83.5 & 97.2 & -- \\
  & Twitter    & 93.5 & 99.5 & -- \\
\midrule
\multirow{4}{*}{BITE}
  & HSOL       & 93.5 & 97.7 & -- \\
  & SST-2      & 91.5 & 78.9 & -- \\
  & Offenseval & 84.2 & 98.1 & -- \\
  & Twitter    & 93.2 & 99.8 & -- \\
\midrule
\multirow{4}{*}{\textbf{DND}}
  & HSOL       & \cellcolor{graybg}\textbf{95.5} & 95.6 & \cellcolor{graybg}\textbf{99.2} \\
  & SST-2      & 91.9 & 93.7 & \cellcolor{graybg}\textbf{98.7} \\
  & Offenseval & \cellcolor{graybg}\textbf{84.6} & 89.3 & \cellcolor{graybg}\textbf{99.8} \\
  & Twitter    & \cellcolor{graybg}\textbf{94.2} & 99.0 & \cellcolor{graybg}\textbf{100} \\
\bottomrule
\multicolumn{5}{l}{\footnotesize \textit{Note:} \textbf{Bold} numbers with \colorbox{graybg}{gray background} denote the best performance.} \\
\multicolumn{5}{l}{\footnotesize ``Benign'' indicates the model trained on clean data.} \\
\end{tabular}
\vspace{-5mm}
\end{table}

\begin{figure}[t!]
    \centering
    \includegraphics[width=0.45\textwidth]{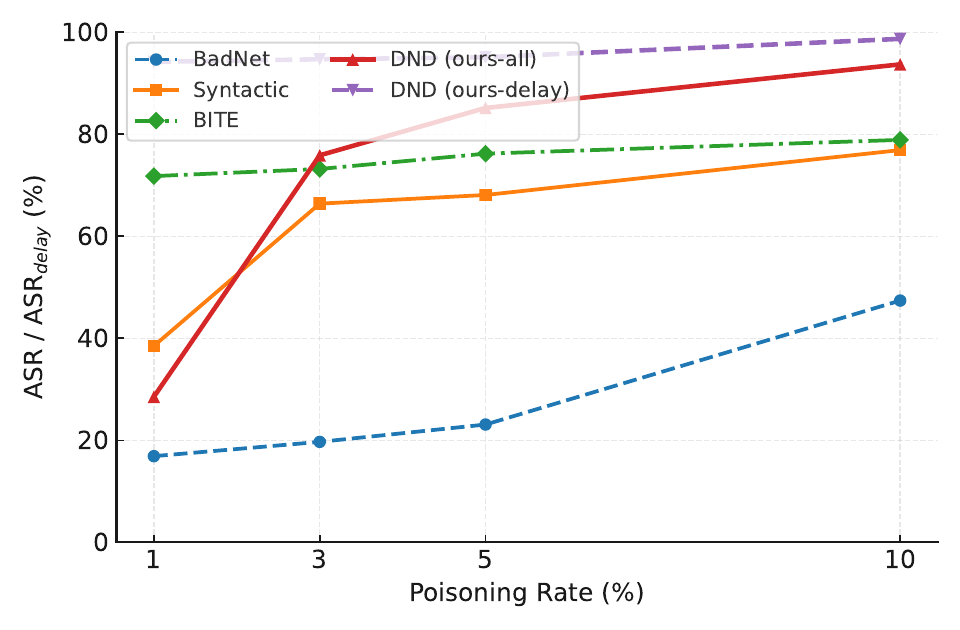}
    \vspace{-3mm}
    \caption{Attack performance vs. poisoning rate on SST-2. DND maintains near-perfect ASR across all ratios, surpassing baselines even with minimal poisoning.}
    \vspace{-7mm}
    \label{fig:rate}
\end{figure}

\paragraph{Effect of Poisoning Rates}
Figure~\ref{fig:rate} presents the attack performance across different poisoning ratios on the SST-2 dataset. 
As the poisoning rate increases, both the ASR of baseline methods and the delayed ASR (ASR$_{delay}$) of DND rise consistently, reflecting the enhanced influence of poisoned samples during training. 
DND (ours-all) surpasses all baselines under every setting, while DND (ours-delay) remains near-saturated (above 94\%) even with only 1\% poisoning. 

These results indicate that the proposed delayed mechanism maintains strong effectiveness under minimal contamination. 
At low poisoning ratios (1\%–3\%), the model benefits from the nonlinear accumulation process, allowing the trigger signal to gradually build up and activate once the internal threshold (Equation~\ref{eq:o}) is reached. 
This behavior demonstrates DND’s practicality in realistic scenarios, where only a small fraction of poisoned data can still result in high attack success rates due to its time-dependent activation dynamics.

\begin{table*}[t]
\centering
\caption{Comparative Performance of DND and Baseline Methods Against State-of-the-Art Defenses (HSOL Dataset).}
\label{tab:defense_comparison}
\renewcommand{\arraystretch}{1.4} 
\small 


\begin{tabular*}{\textwidth}{@{\extracolsep{\fill}} l ccc c ccc c}
\toprule
\multirow{2.5}{*}{\textbf{Defenses}} & \multicolumn{4}{c}{\textbf{CA (\%)}} & \multicolumn{4}{c}{\textbf{ASR / ASR$_{delay}$ (\%)}} \\
\cmidrule(lr){2-5} \cmidrule(lr){6-9}
 & BadNet & Syntactic & BITE & \textbf{DND (Ours)} & BadNet & Syntactic & BITE & \textbf{DND (Ours)} \\
\midrule

Original 
& 95.3 & 93.8 & 93.5 & \cellcolor{graybg}\textbf{95.5} 
& 23.9 & 96.7 & 97.7 & \cellcolor{graybg}\textbf{99.2} \\
\midrule

ONION 
& 95.5 \footnotesize{($\uparrow$0.2)} & 90.2 \footnotesize{($\downarrow$3.6)} & 90.2 \footnotesize{($\downarrow$3.3)} & 95.3 \footnotesize{($\downarrow$0.2)} 
& 19.5 \footnotesize{($\downarrow$4.4)} & \cellcolor{graybg}\textbf{97.1} \footnotesize{($\uparrow$0.4)} & 91.2 \footnotesize{($\downarrow$6.5)} & 96.9 \footnotesize{($\downarrow$2.3)} \\

STRIP 
& 94.7 \footnotesize{($\downarrow$0.6)} & 92.6 \footnotesize{($\uparrow$1.2)} & 91.4 \footnotesize{($\downarrow$2.1)} & 93.9 \footnotesize{($\downarrow$1.6)} 
& 23.9 \footnotesize{($\downarrow$0.0)} & 91.8 \footnotesize{($\downarrow$4.9)} & 94.1 \footnotesize{($\downarrow$3.6)} & \cellcolor{graybg}\textbf{97.3} \footnotesize{($\downarrow$1.9)} \\

RAP 
& 94.0 \footnotesize{($\downarrow$1.3)} & 91.5 \footnotesize{($\downarrow$2.3)} & 90.3 \footnotesize{($\downarrow$3.4)} & 93.1 \footnotesize{($\downarrow$2.4)} 
& 21.1 \footnotesize{($\downarrow$2.8)} & 90.3 \footnotesize{($\downarrow$6.4)} & 94.4 \footnotesize{($\downarrow$3.3)} & \cellcolor{graybg}\textbf{97.8} \footnotesize{($\downarrow$1.4)} \\

CUBE 
& 95.3 \footnotesize{($\uparrow$0.0)} & 93.8 \footnotesize{($\downarrow$0.0)} & 93.4 \footnotesize{($\downarrow$0.1)} & 94.8 \footnotesize{($\downarrow$0.7)} 
& 23.9 \footnotesize{($\uparrow$0.0)} & 96.7 \footnotesize{($\downarrow$0.0)} & 97.7 \footnotesize{($\downarrow$0.0)} & \cellcolor{graybg}\textbf{97.9} \footnotesize{($\downarrow$1.3)} \\

\bottomrule
\multicolumn{9}{l}{\footnotesize \textit{Note:} \textbf{Bold} numbers with \colorbox{graybg}{gray background} denote the best attack performance retained under each defense.} \\
\end{tabular*}
\vspace{-10pt}
\end{table*}

\begin{figure*}[t!]
    \centering
    \includegraphics[scale=0.55]{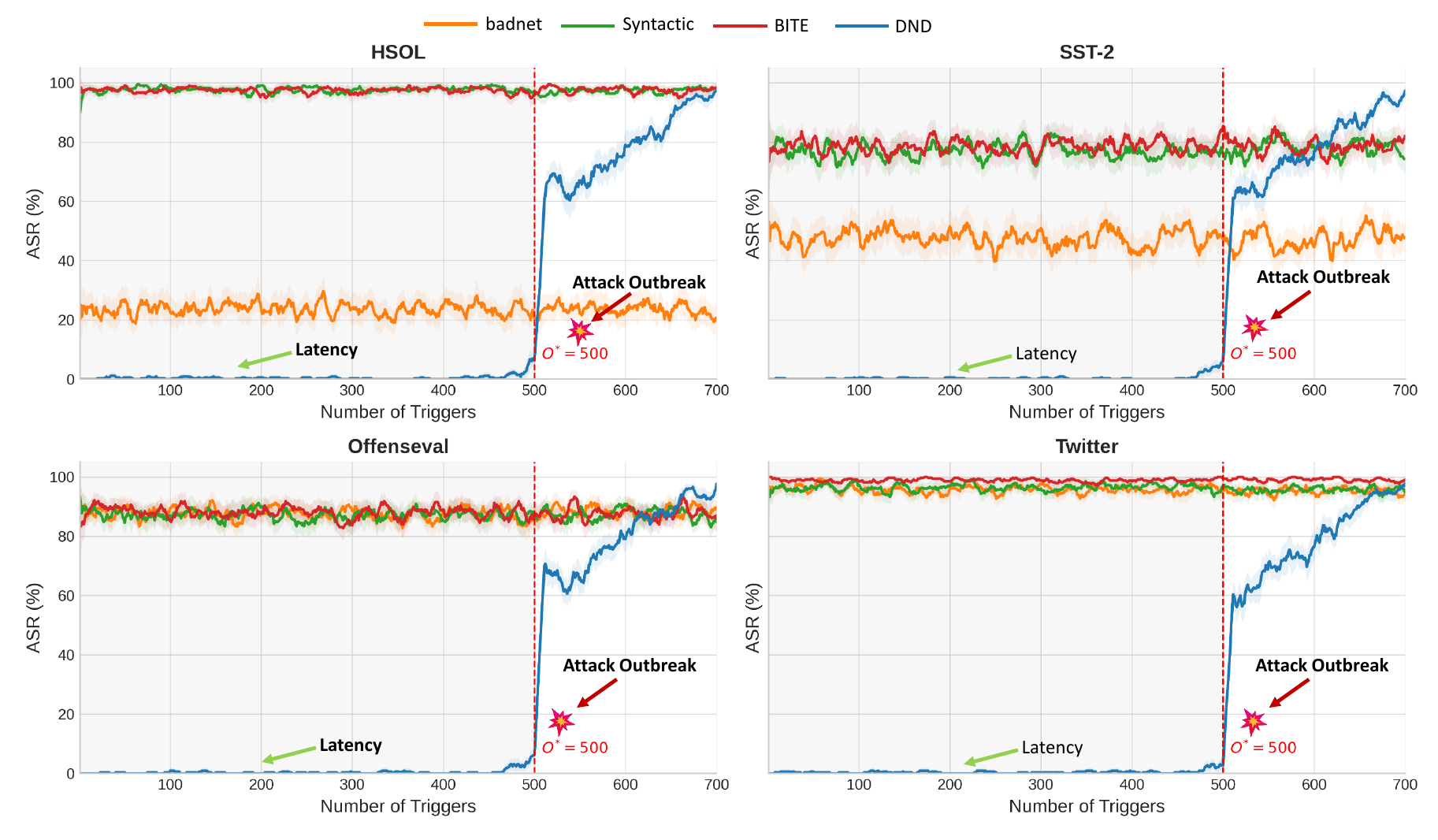}
    \vspace{-3mm}
    \caption{To compare the trigger attack curves of different methods across four datasets, we visualize the moments when the trigger attacks occur. This allows for the observation of the immediacy and latency in backdoor behavior across different methods.}
    \label{fig: attack line}
\end{figure*}

\paragraph{Analysis of delayed attack effect}
To demonstrate the delayed characteristics of our backdoor attack, we set the activation threshold of DND to 500 using a nonlinear decay operator in our experiments. Since the activation threshold was set to 500, we aimed to better highlight the delayed effect by limiting the number of triggers to 700 for these evaluations. This adjustment allows for a clearer distinction between the latent phase and the eventual backdoor activation, thus emphasizing the effectiveness of the delayed activation mechanism. We conducted trigger curve analyses for DND and other baseline methods on four classification task datasets with a poisoning rate of 10$\%$, calculating the attack success probability within a sliding window of 40 triggers. The results are shown in Figure~\ref{fig: attack line}. It can be observed that DND exhibits almost no successful attacks before trigger index 500, with only a few sporadic successes occurring between 400 and 500. This behavior reflects the effect of the nonlinear decay operator, which assigns a gradually increasing trigger probability prior to the outbreak point, allowing for a low but non-zero chance of early activation. Consequently, although the attack is primarily latent during this phase, occasional activations may occur. This indicates a significant delay effect in DND's latency mode. After the number of triggers exceeds 500 (This is consistent with the burst threshold of 500 set in the experiment), the attack success rate rises sharply, reaching nearly 100$\%$ in most cases and marking a clear transition into the outbreak phase, where the method consistently demonstrates strong attack performance—consistent with the intended characteristics of a delayed backdoor.

In contrast, the attack curve of BadNet remains consistently at a relatively low level with visible fluctuations across all trigger events, reflecting limited and unstable attack effectiveness. Syntactic and BITE, on the other hand, maintain consistently high and stable success rates throughout the trigger timeline, without any observable delay or phase shift. These behaviors are indicative of instant-response backdoor mechanisms.
We note that the rapid increase in attack success near the beginning for baseline methods (e.g., Syntactic and BITE), as well as the minor dip in DND's curve around trigger index 700 (), are artifacts introduced by the sliding window-based computation of ASR. Specifically, at the start of the trigger sequence, the window gradually fills up, leading to a progressive increase in the number of samples used for calculation, which causes the ASR to appear to rise. Similarly, localized fluctuations, such as the drop observed near index 700, may occur when the window contains a few unsuccessfully triggered samples. These variations do not reflect inconsistencies or failures in the attack mechanisms themselves, but rather stem from the statistical smoothing behavior of the sliding window approach.


\paragraph{Robustness against defense methods} We evaluated the robustness of our delayed attack against state-of-the-art defenses. To assess this, we conducted experiments on the HSOL dataset with a poisoning rate of 10$\%$, and the results are presented in Table \ref{tab:defense_comparison}. In the experimental results, DND’s ASR$_{delay}$ decreased by 2.3$\%$ against ONION, and by only 1.9$\%$, 1.4$\%$, and 1.3$\%$ against STRIP, RAP, and CUBE, respectively. This indicates that while the four defense methods mitigate DND’s attack to some extent, their effectiveness is very limited, as DND consistently maintains a high ASR$_{delay}$. Additionally, we observe that the baseline methods also demonstrate considerable resistance; however, DND’s robustness remains superior to the baseline methods in most cases.
The defense mechanism of DND relies on its dual modes: the ``latency mode'' and the ``outbreak mode.'' In the latency mode, the model behaves normally, even when the trigger is activated, preventing immediate abnormal behavior. This allows DND to evade detection by mechanisms like ONION and STRIP, which rely on identifying input anomalies or consistency checks. Since the attack does not manifest immediately, these defenses struggle to detect it through simple input perturbations or model behavior monitoring. Similarly, RAP, which adds robustness-aware perturbations to defend against backdoor attacks, is typically more effective against immediate attacks that trigger malicious responses instantly. However, it proves less effective against the delayed activation strategy employed by DND. Moreover, CUBE evaluates the effectiveness of various backdoor defenses through comparative analysis and also has minimal impact when confronting DND. This shows that the defense against delayed backdoor attacks cannot be ignored. We can develop methods to defend against delayed backdoor attacks in the future. 
\subsection{Attack Ablation Experiment}
\label{Ablation Experiment}

\paragraph{Impact of the Hyperparameters}
To examine the influence of the nonlinear decay parameters $a$, $b$, and $c$ (Eq.~\ref{eq:o}) on the delayed activation behavior, we conduct an ablation study on the HSOL dataset. Table~\ref{tab:hyperparams} summarizes the representative results.

\begin{table}[t]
\centering
\caption{Hyperparameter Sensitivity Analysis (HSOL Dataset).}
\label{tab:hyperparams}
\resizebox{\columnwidth}{!}{%
\begin{tabular}{@{}llcc@{}}
\toprule
\textbf{Fixed Parameters} & \textbf{Varying Parameter} & \textbf{CA (\%)} & \textbf{ASR$_{delay}$ (\%)} \\ \midrule
\multirow{2}{*}{$b=2, c=500$} & $a = 1 \times 10^5$ & 94.91 & 99.2 \\
 & $a = 5 \times 10^5$ & 94.72 & 99.2 \\ \addlinespace[3pt]
\multirow{2}{*}{$a=2.5 \times 10^5, c=500$} & $b = 1.5$ & 94.65 & 99.2 \\
 & $b = 2.5$ & 94.93 & 99.2 \\ \addlinespace[3pt]
\multirow{2}{*}{$a=2.5 \times 10^5, b=2$} & $c = 300$ & 94.93 & 99.2 \\
 & $c = 700$ & 94.61 & 99.2 \\ \bottomrule
\end{tabular}%
}
\vspace{-10pt}
\end{table}

\begin{table}[t]
\centering
\caption{Comprehensive Analysis: Attack Ablation and Defense Robustness.}
\label{tab:master_ablation}
\resizebox{\columnwidth}{!}{%
\begin{tabular}{@{}llcc@{}}
\toprule
\textbf{Experimental Setting} & \textbf{Parameter / Condition} & \textbf{CA (\%)} & \textbf{ASR$_{delay}$ (\%)} \\ \midrule
\multicolumn{4}{c}{\cellcolor{gray!10}\textbf{Part I: Attack Configuration Ablation}} \\ \midrule
\multirow{3}{*}{Dataset Size} & 10,000 & 92.1 & 99.1 \\
 & 40,000 & 92.0 & 99.1 \\
 & 70,042 (Original) & 91.9 & 98.7 \\ \midrule
\multirow{3}{*}{Count Condition} & 1 & 95.8 & 98.8 \\
 & 100 & 95.8 & 99.0 \\
 & 500 & 95.5 & 99.2 \\ \midrule
\multirow{4}{*}{Poisoned Sample Count} & 100 & 96.3 & 0.0 \\
 & 400 & 96.2 & 0.0 \\
 & 3,000 & 95.7 & 99.2 \\
 & 7,000 & 95.5 & 99.2 \\ \midrule
\multicolumn{4}{c}{\cellcolor{gray!10}\textbf{Part II: Defense Robustness Ablation}} \\ \midrule
\multirow{3}{*}{Fine-Pruning (Rate)} & 0.01 & 94.6 & 99.2 \\
 & 0.05 & 94.3 & 99.2 \\
 & 0.10 & 94.2 & 99.2 \\ \midrule
\multirow{4}{*}{MDP ($\epsilon$)} & 0.01 & 94.6 & 99.2 \\
 & 0.05 & 94.6 & 99.2 \\
 & 0.10 & 94.5 & 99.2 \\
 & 0.20 & 94.4 & 99.2 \\ \bottomrule
\end{tabular}%
}
\vspace{-10pt}
\end{table}

The results indicate that ASR$_{delay}$ remains consistently saturated ($\approx$ 99.2\%) across all configurations, while CA fluctuates only within a narrow range ($\pm$ 0.3\%). This observation suggests that the hyperparameters mainly regulate the activation threshold rather than altering the final attack success or benign performance. Specifically, parameter $a$ controls the initial trigger magnitude, $b$ determines the decay rate, and $c$ defines the activation threshold. Together, these parameters determine the minimal activation step described in Eq.~\ref{eq:o}.

\begin{figure}[t]
  \centering
  \begin{minipage}{0.49\columnwidth}
    \centering
    \includegraphics[width=\linewidth]{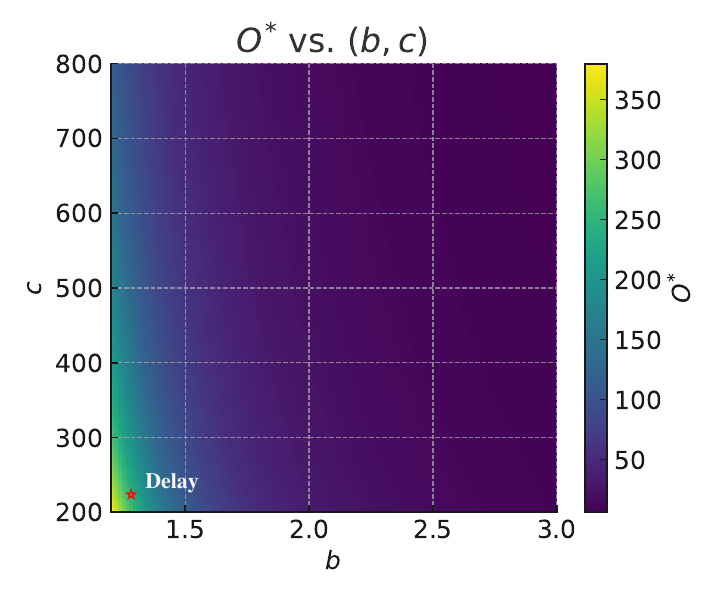}
  \end{minipage}
  \hfill 
  \begin{minipage}{0.49\columnwidth}
    \centering
    \includegraphics[width=\linewidth]{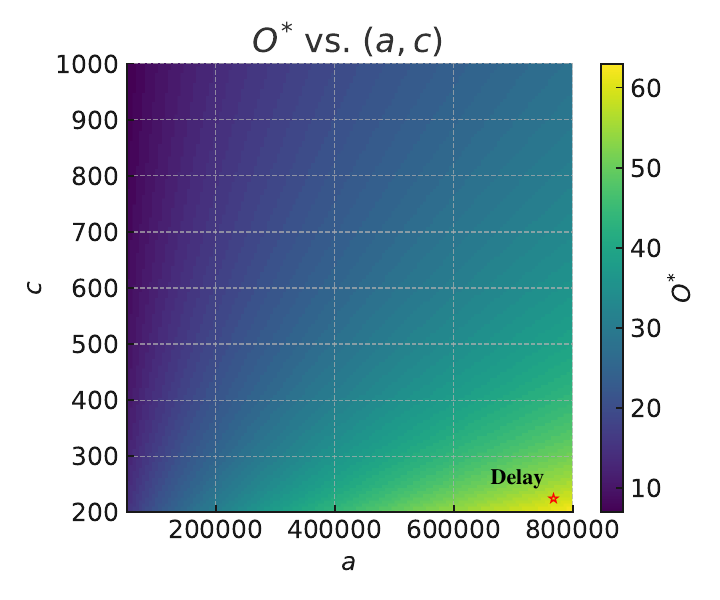}
  \end{minipage}
  
  \vspace{-3mm} 
  \caption{Hyperparameter sensitivity analysis. Left: $O^{*}$ vs.\ $(b,c)$, showing larger $b$ or $c$ shortens latency. Right: $O^{*}$ vs.\ $(a,c)$, showing higher $a$ or lower $c$ extends latency.}
  \label{fig:sensitivity_analysis}
  \vspace{-5mm} 
\end{figure}

As illustrated in Figure~\ref{fig:sensitivity_analysis}, the surface of $O^{*}$ exhibits smooth and monotonic variations. Figure~\ref{fig:sensitivity_analysis} (left) shows that increasing $b$ or $c$ leads to faster accumulation and shorter latency, while Figure~\ref{fig:sensitivity_analysis}  (right) reveals that higher $a$ or lower $c$ prolongs the delay window. Such consistent gradients confirm that the nonlinear decay function provides an interpretable mechanism for adjusting the activation latency without affecting the final ASR$_{delay}$. This controllability allows the adversary to fine-tune the ``patience'' of the backdoor by modulating $(a,b,c)$ alone. From a defense perspective, these findings highlight a critical insight: post-activation success metrics alone are insufficient for detection. Instead, robust defense mechanisms should incorporate \textit{temporal profiling} of latent activation dynamics.
\paragraph{Scalability and Adaptability Analysis}
We conducted a comprehensive ablation study to evaluate DND's sensitivity to three key experimental conditions, as summarized in {Table \ref{tab:master_ablation} (Part I).
First, to assess \textit{scalability}, we varied the dataset size from 10,000 to the full 70,042 samples. Results show negligible fluctuations in ASR ($<0.4\%$) and stable FRR, confirming DND's stability across data scales.
Second, we examined the \textit{Count Condition}. The attack remains robust (ASR $>98\%$) regardless of whether the threshold is set to 1, 100, or 500, demonstrating reliable activation control.
Finally, we analyzed the impact of \textit{poisoned sample counts}. We observe a sharp phase transition: the attack remains dormant (0\% ASR) when poisoned samples are insufficient to reach the activation threshold (e.g., $<500$), but achieves near-perfect efficacy ($>99\%$) once the threshold is crossed. This confirms that the effectiveness of DND relies on the cumulative state of the triggers and the delayed activation mechanism.
\subsection{Defense Ablation Experiment}
\label{defense}
We further evaluated the robustness of DND against two structure-aware mitigation strategies, detailed in Table \ref{tab:master_ablation} (Part II).
\textit{Fine-pruning} attempts to remove compromised neurons by pruning less active connections. Even at a pruning rate of 0.1 (removing 10\% of neurons), DND maintains an ASR of 99.2\%, indicating that the injected logic is not isolated to sparse, easily pruned connections.

Similarly, \textit{Masking-Differential Prompting (MDP)} applies input perturbations (controlled by $\epsilon$-MDP) to disrupt trigger patterns. DND exhibits high resilience with stable ASR across varying $\epsilon$ values (0.01 to 0.2). This resilience is attributed to the delayed activation mechanism, which decouples the trigger and activation process, making it less sensitive to local input perturbations than traditional static triggers. Since the backdoor attack is not solely dependent on specific neurons but on a temporal delay, its effectiveness remains intact even when the model is subjected to MDP perturbations.
\section{Discussion}
\label{sec:discussion}


\subsection{Implications and Evaluation Challenges}
\label{sec:Evaluation}
As noted in the Introduction, a key implication of the DBA paradigm is its ability to weaponize common, high-frequency words as triggers. This subsection further examines that idea: we contend that such triggers represent a more powerful and more covert form of backdoor. An instant-activation backdoor that uses a ubiquitous token (e.g., ``the'') is practically infeasible because it would catastrophically degrade the model’s clean accuracy. Consequently, the delayed-activation mechanism is not merely an enhancement for stealth, but the enabling mechanism that makes this class of attacks viable.

Moreover, this reveals a second fundamental blind spot, this time in evaluation paradigms. The standard definitions of CA and ASR are implicitly grounded in the assumption of a statistically separable, low-frequency trigger. When the trigger becomes a common, high-frequency word, these metrics collapse, as the distinction between ``clean'' and ``poisoned'' test sets becomes meaningless. This highlights that our work not only introduces a new attack vector, but also calls for the development of more advanced and context-aware evaluation frameworks. Future metrics may need to be stateful, capturing not only single-instance accuracy but also a model’s behavioral consistency and temporal stability across long-term interaction histories.

\subsection{Limitations and Future Work}
\label{sec:limitations_future_work}
Although our DND prototype provides a successful proof of concept for the DBA paradigm, it should be regarded as an early-stage exploration. Its current limitations highlight several promising avenues for future investigation.

\paragraph{Limitations}
Our current implementation relies on explicit structural modifications. Although these modifications are difficult to detect by automated static analysis, they may remain vulnerable to rigorous manual white-box testing. Similar limitations have also been observed in prior studies on structural or architectural backdoors, which noted that manual inspection of model components can still reveal the embedded malicious logic~\cite{bober2023architectural,langford2025architectural,liu2022loneneuron}. Future work should explore more deeply obfuscated or dynamically constructed architectural backdoors that resist human inspection.

\paragraph{Future Work}
We advocate for the development of a new class of stateful defense mechanisms, which possess memory capabilities and can analyze model behavior over extended temporal windows. Promising directions for future research include:
(i) Structural integrity verification: developing techniques to construct ``clean'' architectural baselines and detect unauthorized stateful modifications to model logic;
(ii) Latent representation monitoring: moving beyond single-input outlier detection to track statistical drifts in latent representations across long input sequences; and
(iii) Runtime behavioral analysis: recording and analyzing high-level behavioral patterns of pre-trained models, such as prediction sequences and temporal variations in confidence, to identify deviations from expected operational dynamics.

\section{Conclusion}
\label{sec:conclusion}
In this paper, we have revisited the foundational ``immediacy assumption'' in backdoor research and introduced DBA, a novel threat paradigm in which malicious activation is temporally decoupled from trigger exposure. We have demonstrated that this temporal dimension constitutes a viable yet largely unprotected attack surface in PTMs. To substantiate its practical feasibility, we have designed DND, a proof-of-concept prototype that implements a stateful and controllable delayed activation mechanism. Comprehensive experiments, evaluated through a dual-metric framework, have provided the first empirical evidence of a successful and precisely controlled delay process. The results have shown that our prototype can remain dormant for a parameterized duration, bypass multiple state-of-the-art defenses, and later unleash a highly effective attack, all while maintaining model performance on clean data. Although DND represents an initial exploration and has its own limitations, the findings highlight an important implication: the security community must look beyond instantaneous anomaly detection. Future research should focus on developing stateful, temporal-aware, and adaptive defense mechanisms to counter this emerging class of patient and strategically timed threats.



\bibliographystyle{IEEEtran}
\bibliography{example_paper.bib}

@article{gan2021triggerless,
  title={Triggerless backdoor attack for NLP tasks with clean labels},
  author={Gan, Leilei and Li, Jiwei and Zhang, Tianwei and Li, Xiaoya and Meng, Yuxian and Wu, Fei and Yang, Yi and Guo, Shangwei and Fan, Chun},
  journal={arXiv preprint arXiv:2111.07970},
  year={2021}
}

@article{li2021backdoor,
  title={Backdoor attacks on pre-trained models by layerwise weight poisoning},
  author={Li, Linyang and Song, Demin and Li, Xiaonan and Zeng, Jiehang and Ma, Ruotian and Qiu, Xipeng},
  journal={arXiv preprint arXiv:2108.13888},
  year={2021}
}

@article{yan2022bite,
  title={Bite: Textual backdoor attacks with iterative trigger injection},
  author={Yan, Jun and Gupta, Vansh and Ren, Xiang},
  journal={arXiv preprint arXiv:2205.12700},
  year={2022}
}

@inproceedings{pan2022hidden,
  title={Hidden trigger backdoor attack on $\{$NLP$\}$ models via linguistic style manipulation},
  author={Pan, Xudong and Zhang, Mi and Sheng, Beina and Zhu, Jiaming and Yang, Min},
  booktitle={31st USENIX Security Symposium (USENIX Security 22)},
  pages={3611--3628},
  year={2022}
}

@inproceedings{chen2021badnl,
  title={Badnl: Backdoor attacks against nlp models with semantic-preserving improvements},
  author={Chen, Xiaoyi and Salem, Ahmed and Chen, Dingfan and Backes, Michael and Ma, Shiqing and Shen, Qingni and Wu, Zhonghai and Zhang, Yang},
  booktitle={Proceedings of the 37th Annual Computer Security Applications Conference},
  pages={554--569},
  year={2021}
}

@inproceedings{li2021hidden,
  title={Hidden backdoors in human-centric language models},
  author={Li, Shaofeng and Liu, Hui and Dong, Tian and Zhao, Benjamin Zi Hao and Xue, Minhui and Zhu, Haojin and Lu, Jialiang},
  booktitle={Proceedings of the 2021 ACM SIGSAC Conference on Computer and Communications Security},
  pages={3123--3140},
  year={2021}
}

@inproceedings{liu2017neural,
  title={Neural trojans},
  author={Liu, Yuntao and Xie, Yang and Srivastava, Ankur},
  booktitle={2017 IEEE International Conference on Computer Design (ICCD)},
  pages={45--48},
  year={2017},
  organization={IEEE}
}

@inproceedings{sheng2023punctuation,
  title={Punctuation Matters! Stealthy Backdoor Attack for Language Models},
  author={Sheng, Xuan and Li, Zhicheng and Han, Zhaoyang and Chang, Xiangmao and Li, Piji},
  booktitle={CCF International Conference on Natural Language Processing and Chinese Computing},
  pages={524--536},
  year={2023},
  organization={Springer}
}

@article{kurita2020weight,
  title={Weight poisoning attacks on pre-trained models},
  author={Kurita, Keita and Michel, Paul and Neubig, Graham},
  journal={arXiv preprint arXiv:2004.06660},
  year={2020}
}

@inproceedings{socher2013recursive,
  title={Recursive deep models for semantic compositionality over a sentiment treebank},
  author={Socher, Richard and Perelygin, Alex and Wu, Jean and Chuang, Jason and Manning, Christopher D and Ng, Andrew Y and Potts, Christopher},
  booktitle={Proceedings of the 2013 conference on empirical methods in natural language processing},
  pages={1631--1642},
  year={2013}
}

@book{cover1999elements,
  title={Elements of information theory},
  author={Cover, Thomas M},
  year={1999},
  publisher={John Wiley \& Sons}
}

@book{pinsker1960information,
  title={Information and Information Stability of Random Variables and Processes},
  author={Pinsker, M. S.},
  year={1960},
  publisher={Holden-Day},
  note={Translated and edited by A. Feinstein}
}

@article{sason2016fdivergence,
  title={f-Divergence Inequalities},
  author={Sason, Igal and Verd{\'u}, Sergio},
  journal={IEEE Transactions on Information Theory},
  volume={62},
  number={11},
  pages={5973--6006},
  year={2016},
  doi={10.1109/TIT.2016.2603152}
}

@inproceedings{davidson2017automated,
  title={Automated hate speech detection and the problem of offensive language},
  author={Davidson, Thomas and Warmsley, Dana and Macy, Michael and Weber, Ingmar},
  booktitle={Proceedings of the international AAAI conference on web and social media},
  volume={11},
  number={1},
  pages={512--515},
  year={2017}
}

@article{zampieri2019predicting,
  title={Predicting the type and target of offensive posts in social media},
  author={Zampieri, Marcos and Malmasi, Shervin and Nakov, Preslav and Rosenthal, Sara and Farra, Noura and Kumar, Ritesh},
  journal={arXiv preprint arXiv:1902.09666},
  year={2019}
}

@article{gu2017badnets,
  title={Badnets: Identifying vulnerabilities in the machine learning model supply chain},
  author={Gu, Tianyu and Dolan-Gavitt, Brendan and Garg, Siddharth},
  journal={arXiv preprint arXiv:1708.06733},
  year={2017}
}

@inproceedings{founta2018large,
  title={Large scale crowdsourcing and characterization of twitter abusive behavior},
  author={Founta, Antigoni Maria and Djouvas, Constantinos and Chatzakou, Despoina and Leontiadis, Ilias and Blackburn, Jeremy and Stringhini, Gianluca and Vakali, Athena I and Sirivianos, Michael and Kourtellis, Nicolas},
  booktitle={International AAAI Conference on Web and Social Media},
  year={2018}
}

@inproceedings{liu2018fine,
  title={Fine-pruning: Defending against backdooring attacks on deep neural networks},
  author={Liu, Kang and Dolan-Gavitt, Brendan and Garg, Siddharth},
  booktitle={International symposium on research in attacks, intrusions, and defenses},
  pages={273--294},
  year={2018},
  organization={Springer}
}

@article{qi2020onion,
  title={Onion: A simple and effective defense against textual backdoor attacks},
  author={Qi, Fanchao and Chen, Yangyi and Li, Mukai and Yao, Yuan and Liu, Zhiyuan and Sun, Maosong},
  journal={arXiv preprint arXiv:2011.10369},
  year={2020}
}

@article{cui2022unified,
  title={A unified evaluation of textual backdoor learning: Frameworks and benchmarks},
  author={Cui, Ganqu and Yuan, Lifan and He, Bingxiang and Chen, Yangyi and Liu, Zhiyuan and Sun, Maosong},
  journal={Advances in Neural Information Processing Systems},
  volume={35},
  pages={5009--5023},
  year={2022}
}

@article{gao2021design,
  title={Design and evaluation of a multi-domain trojan detection method on deep neural networks},
  author={Gao, Yansong and Kim, Yeonjae and Doan, Bao Gia and Zhang, Zhi and Zhang, Gongxuan and Nepal, Surya and Ranasinghe, Damith C and Kim, Hyoungshick},
  journal={IEEE Transactions on Dependable and Secure Computing},
  volume={19},
  number={4},
  pages={2349--2364},
  year={2021},
  publisher={IEEE}
}

@article{yang2021rap,
  title={Rap: Robustness-aware perturbations for defending against backdoor attacks on nlp models},
  author={Yang, Wenkai and Lin, Yankai and Li, Peng and Zhou, Jie and Sun, Xu},
  journal={arXiv preprint arXiv:2110.07831},
  year={2021}
}

@inproceedings{qi2021hidden,
  title={Hidden Killer: Invisible Textual Backdoor Attacks with Syntactic Trigger},
  author={Qi, Fanchao and Li, Mukai and Chen, Yangyi and Zhang, Zhengyan and Liu, Zhiyuan and Wang, Yasheng and Sun, Maosong},
  booktitle={Proceedings of the 59th Annual Meeting of the Association for Computational Linguistics and the 11th International Joint Conference on Natural Language Processing (Volume 1: Long Papers)},
  pages={443--453},
  year={2021}
}

@inproceedings{wang2019neural,
  title={Neural cleanse: Identifying and mitigating backdoor attacks in neural networks},
  author={Wang, Bolun and Yao, Yuanshun and Shan, Shawn and Li, Huiying and Viswanath, Bimal and Zheng, Haitao and Zhao, Ben Y},
  booktitle={2019 IEEE symposium on security and privacy (SP)},
  pages={707--723},
  year={2019},
  organization={IEEE}
}

@inproceedings{li2021deeppayload,
  title={Deeppayload: Black-box backdoor attack on deep learning models through neural payload injection},
  author={Li, Yuanchun and Hua, Jiayi and Wang, Haoyu and Chen, Chunyang and Liu, Yunxin},
  booktitle={2021 IEEE/ACM 43rd International Conference on Software Engineering (ICSE)},
  pages={263--274},
  year={2021},
  organization={IEEE}
}

@inproceedings{liu2022loneneuron,
  title={LoneNeuron: a highly-effective feature-domain neural trojan using invisible and polymorphic watermarks},
  author={Liu, Zeyan and Li, Fengjun and Li, Zhu and Luo, Bo},
  booktitle={Proceedings of the 2022 ACM SIGSAC Conference on Computer and Communications Security},
  pages={2129--2143},
  year={2022}
}

@inproceedings{tang2020embarrassingly,
  title={An embarrassingly simple approach for trojan attack in deep neural networks},
  author={Tang, Ruixiang and Du, Mengnan and Liu, Ninghao and Yang, Fan and Hu, Xia},
  booktitle={Proceedings of the 26th ACM SIGKDD international conference on knowledge discovery \& data mining},
  pages={218--228},
  year={2020}
}

@article{patel2025towards,
  title={Towards Secure MLOps: Surveying Attacks, Mitigation Strategies, and Research Challenges},
  author={Patel, Raj and Tripathi, Himanshu and Stone, Jasper and Golilarz, Noorbakhsh Amiri and Mittal, Sudip and Rahimi, Shahram and Chaudhary, Vini},
  journal={arXiv preprint arXiv:2506.02032},
  year={2025}
}

@article{hu2025understanding,
  title={Understanding Large Language Model Supply Chain: Structure, Domain, and Vulnerabilities},
  author={Hu, Yanzhe and Wang, Shenao and Nie, Tianyuan and Zhao, Yanjie and Wang, Haoyu},
  journal={arXiv preprint arXiv:2504.20763},
  year={2025}
}

@article{li2023trustworthy,
  title={Trustworthy AI: From principles to practices},
  author={Li, Bo and Qi, Peng and Liu, Bo and Di, Shuai and Liu, Jingen and Pei, Jiquan and Yi, Jinfeng and Zhou, Bowen},
  journal={ACM Computing Surveys},
  volume={55},
  number={9},
  pages={1--46},
  year={2023},
  publisher={ACM New York, NY}
}

@article{kaur2022trustworthy,
  title={Trustworthy artificial intelligence: a review},
  author={Kaur, Davinder and Uslu, Suleyman and Rittichier, Kaley J and Durresi, Arjan},
  journal={ACM computing surveys (CSUR)},
  volume={55},
  number={2},
  pages={1--38},
  year={2022},
  publisher={ACM New York, NY}
}

@inproceedings{cordeiro2025neural,
  title={Neural network verification is a programming language challenge},
  author={Cordeiro, Lucas C and Daggitt, Matthew L and Girard-Satabin, Julien and Isac, Omri and Johnson, Taylor T and Katz, Guy and Komendantskaya, Ekaterina and Lemesle, Augustin and Manino, Edoardo and {\v{S}}inkarovs, Artjoms and others},
  booktitle={European Symposium on Programming},
  pages={206--235},
  year={2025},
  organization={Springer}
}

@article{sbai2025model,
  title={Model checking deep neural networks: opportunities and challenges},
  author={Sbai, Zohra},
  journal={Frontiers in Computer Science},
  volume={7},
  pages={1557977},
  year={2025},
  publisher={Frontiers Media SA}
}

@article{bai2024backdoor,
  title={Backdoor attack and defense on deep learning: A survey},
  author={Bai, Yang and Xing, Gaojie and Wu, Hongyan and Rao, Zhihong and Ma, Chuan and Wang, Shiping and Liu, Xiaolei and Zhou, Yimin and Tang, Jiajia and Huang, Kaijun and others},
  journal={IEEE Transactions on Computational Social Systems},
  year={2024},
  publisher={IEEE}
}

@article{li2022backdoor,
  title={Backdoor learning: A survey},
  author={Li, Yiming and Jiang, Yong and Li, Zhifeng and Xia, Shu-Tao},
  journal={IEEE transactions on neural networks and learning systems},
  volume={35},
  number={1},
  pages={5--22},
  year={2022},
  publisher={IEEE}
}

@article{brown2020language,
  title={Language models are few-shot learners},
  author={Brown, Tom and Mann, Benjamin and Ryder, Nick and Subbiah, Melanie and Kaplan, Jared D and Dhariwal, Prafulla and Neelakantan, Arvind and Shyam, Pranav and Sastry, Girish and Askell, Amanda and others},
  journal={Advances in neural information processing systems},
  volume={33},
  pages={1877--1901},
  year={2020}
}

@inproceedings{langford2025architectural,
  title={Architectural neural backdoors from first principles},
  author={Langford, Harry and Shumailov, Ilia and Zhao, Yiren and Mullins, Robert and Papernot, Nicolas},
  booktitle={2025 IEEE Symposium on Security and Privacy (SP)},
  pages={1657--1675},
  year={2025},
  organization={IEEE}
}

@inproceedings{yang2021rethinking,
  title={Rethinking stealthiness of backdoor attacks against NLP models},
  author={Yang, Jianbo and Jiang, Yuxuan and Cao, Qipeng and Huang, Zhenhua and Yang, Chengwei},
  booktitle={Proceedings of the 59th Annual Meeting of the Association for Computational Linguistics (ACL)},
  pages={554--565},
  year={2021}
}

@inproceedings{yin2025shadow,
  title={Shadow-Activated Backdoor Attacks on Multimodal Large Language Models},
  author={Yin, Ziyi and Ye, Muchao and Cao, Yuanpu and Wang, Jiaqi and Chang, Aofei and Liu, Han and Chen, Jinghui and Wang, Ting and Ma, Fenglong},
  booktitle={Findings of the Association for Computational Linguistics: ACL 2025},
  pages={4808--4829},
  year={2025}
}

@inproceedings{bober2023architectural,
  title={Architectural backdoors in neural networks},
  author={Bober-Irizar, Mikel and Shumailov, Ilia and Zhao, Yiren and Mullins, Robert and Papernot, Nicolas},
  booktitle={Proceedings of the IEEE/CVF Conference on Computer Vision and Pattern Recognition},
  pages={24595--24604},
  year={2023}
}

@inproceedings{tran2018spectral,
  title={Spectral signatures in backdoor attacks},
  author={Tran, Brandon and Li, Jerry and Madry, Aleksander},
  booktitle={Advances in Neural Information Processing Systems (NeurIPS)},
  pages={8011--8021},
  year={2018}
}

@inproceedings{kolouri2020universal,
  title={Universal litmus patterns: Revealing backdoor attacks in CNNs},
  author={Kolouri, Soheil and Saha, Anirban and Pirsiavash, Hamed and Hoffmann, Heiko},
  booktitle={Proceedings of the IEEE/CVF Conference on Computer Vision and Pattern Recognition (CVPR)},
  pages={301--310},
  year={2020}
}

@inproceedings{li2021anti,
  title={Anti-backdoor learning: Training clean models on poisoned data},
  author={Li, Yiming and Li, Baoyuan and Wang, Tao and Lyu, Siwei and Zhang, Yisen},
  booktitle={Advances in Neural Information Processing Systems (NeurIPS)},
  pages={14900--14912},
  year={2021}
}

@article{zheng2021topological,
  title={Topological detection of trojaned neural networks},
  author={Zheng, Songzhu and Zhang, Yikai and Wagner, Hubert and Goswami, Mayank and Chen, Chao},
  journal={Advances in Neural Information Processing Systems},
  volume={34},
  pages={17258--17272},
  year={2021}
}

@article{das2025unmasking,
  title={Unmasking Backdoors: An Explainable Defense via Gradient-Attention Anomaly Scoring for Pre-trained Language Models},
  author={Das, Anindya Sundar and Chen, Kangjie and Bhuyan, Monowar},
  journal={arXiv preprint arXiv:2510.04347},
  year={2025}
}

@inproceedings{kumari2023baybfed,
  title={Baybfed: Bayesian backdoor defense for federated learning},
  author={Kumari, Kavita and Rieger, Phillip and Fereidooni, Hossein and Jadliwala, Murtuza and Sadeghi, Ahmad-Reza},
  booktitle={2023 IEEE symposium on security and privacy (SP)},
  pages={737--754},
  year={2023},
  organization={IEEE}
}

@inproceedings{zhang2023backdoor,
  title={Backdoor defense via deconfounded representation learning},
  author={Zhang, Zaixi and Liu, Qi and Wang, Zhicai and Lu, Zepu and Hu, Qingyong},
  booktitle={Proceedings of the IEEE/CVF Conference on Computer Vision and Pattern Recognition},
  pages={12228--12238},
  year={2023}
}

@article{alam2025reveil,
  title={Reveil: Unconstrained concealed backdoor attack on deep neural networks using machine unlearning},
  author={Alam, Manaar and Lamri, Hithem and Maniatakos, Michail},
  journal={arXiv preprint arXiv:2502.11687},
  year={2025}
}

@inproceedings{chen2025refine,
  title={REFINE: Inversion-Free Backdoor Defense via Model Reprogramming},
  author={Chen, Jiale and Zhang, Ming and Li, Zhi and Xu, Shouling},
  booktitle={International Conference on Learning Representations (ICLR)},
  year={2025}
}

@inproceedings{yuan2025mergehijacking,
  title={Merge Hijacking: Backdoor Attacks to Model Merging of Large Language Models},
  author={Yuan, Zenghui and Xu, Yangming and Shi, Jiawen and Zhou, Pan and Sun, Lichao},
  booktitle={Proceedings of the 63rd Annual Meeting of the Association for Computational Linguistics (ACL)},
  pages={32688--32703},
  year={2025}
}

@article{ning2025badrec,
  title={Exploring Backdoor Attack and Defense for LLM-empowered Recommendations},
  author={Ning, Liangbo and Fan, Wenqi and Li, Qing},
  journal={arXiv preprint arXiv:2504.11182},
  year={2025}
}

@inproceedings{sheng2022survey,
  title={A survey on backdoor attack and defense in natural language processing},
  author={Sheng, Xuan and Han, Zhaoyang and Li, Piji and Chang, Xiangmao},
  booktitle={2022 IEEE 22nd International Conference on Software Quality, Reliability and Security (QRS)},
  pages={809--820},
  year={2022},
  organization={IEEE}
}

@article{shin2024unlearn,
  title={Unlearn to Relearn Backdoors: Deferred Backdoor Functionality Attacks on Deep Learning Models},
  author={Shin, Jeongjin and Park, Sangdon},
  journal={arXiv preprint arXiv:2411.14449},
  year={2024}
}

@inproceedings{tsutsui2023poison,
  title={Poison Egg: Scrambling Federated Learning with Delayed Backdoor Attack},
  author={Tsutsui, Masayoshi and Kaneko, Tatsuya and Takamaeda-Yamazaki, Shinya},
  booktitle={International Conference on Ubiquitous Security},
  pages={191--204},
  year={2023},
  organization={Springer}
}

\end{document}